\documentclass{article}

\usepackage{PRIMEarxiv}

\usepackage[utf8]{inputenc} % allow utf-8 input
\usepackage[T1]{fontenc}    % use 8-bit T1 fonts
\usepackage{hyperref}       % hyperlinks
\usepackage{url}            % simple URL typesetting
\usepackage{booktabs}       % professional-quality tables
\usepackage{amsfonts}       % blackboard math symbols
\usepackage{amsmath}        % more symbols
\usepackage{nicefrac}       % compact symbols for 1/2, etc.
\usepackage{microtype}      % microtypography
\usepackage{float}
\usepackage{lipsum}
\usepackage{fancyhdr}       % header
\usepackage{graphicx}       % graphics
\graphicspath{{media/}}     % organize your images and other figures under media/ folder
\usepackage{amsmath}

\title{How effective is multifactor authentication at deterring cyberattacks?
}

\author{
  Lucas Augusto Meyer \\
  AI for Good Lab \\
  Microsoft Corporation \\
  \texttt{lucas.meyer@microsoft.com} \\
   \And
  Sergio Romero \\
  Identity Security \\
  Microsoft Corporation\\
  \texttt{sezambra@microsoft.com} \\
   \And
  Gabriele Bertoli \\
  Identity Security \\
  Microsoft Corporation\\
  \texttt{gbertoli@microsoft.com} \\
   \And
  Tom Burt \\
  Customer Security and Trust \\
  Microsoft Corporation\\
  \texttt{tburt@microsoft.com} \\
   \And   
  Alex Weinert \\
  Identity Security \\
  Microsoft Corporation\\
  \texttt{alexwe@microsoft.com} \\
   \And
  Juan Lavista Ferres \\
  AI for Good Lab \\
  Microsoft Corporation\\
  \texttt{jlavista@microsoft.com} \\
}

\begin{document}
\maketitle

\begin{abstract}
This study investigates the effectiveness of multifactor authentication (MFA) in protecting commercial accounts from unauthorized access, with an additional focus on accounts with known credential leaks. We employ the benchmark-multiplier method, coupled with manual account review, to evaluate the security performance of various MFA methods in a large dataset of Microsoft Azure Active Directory users exhibiting suspicious activity. Our findings reveal that MFA implementation offers outstanding protection, with over 99.99\% of MFA-enabled accounts remaining secure during the investigation period. Moreover, MFA reduces the risk of compromise by 99.22\% across the entire population and by 98.56\% in cases of leaked credentials. We further demonstrate that dedicated MFA applications, such as Microsoft Authenticator, outperform SMS-based authentication, though both methods provide significantly enhanced security compared to not using MFA. Based on these results, we strongly advocate for the default implementation of MFA in commercial accounts to increase security and mitigate unauthorized access risks.
\end{abstract}

\section{Introduction}

In the past decade, prominent identity providers such as Microsoft, Google and Okta have increasingly adopted risk-based authentication, also known as \textit{challenges}, to enhance security against unauthorized access. These challenges utilize various passive signals to identify anomalous login attempts, including IP geolocation, device and IP address reputation, and the interval between login attempts. Upon detecting irregular login patterns or receiving a user's request to change their password, identity providers issue a challenge requesting supplementary forms of authentication to grant access to the protected resource\cite{freemanWhoAreYou2016}.

Supplementary verification methods can be classified into three categories, also called \textit{factors}: knowledge (something the user \textbf{knows}), possession (something the user \textbf{has}), or inherence (something the user \textbf{is}). When an authentication scheme requires a secondary factor of authentication, it is referred to as two-factor authentication (2FA). More broadly, multifactor authentication (MFA) encompasses authentication methods that require users to present two or more factors to the authentication mechanism \cite{ogormanComparingPasswordsTokens2003}. 

Although there is a lot of variability on factors required to authenticate consumer accounts, companies such as Microsoft and Okta that provide authentication services to enterprises primarily require a possession verification method, sending a code to a device that the user possesses\cite{liveretosCustomerIdentityAccess2022}. Various methods exist for code generation and delivery, including SMS, dedicated mobile applications like Microsoft Authenticator, or authentication-specific devices such as Yubikey \cite{dasWhyJohnnyDoesn2018a}. To use these supplementary authentication measures, users must pre-register them with their accounts. However, the increased friction of pre-registering and frequently verifying a code on a secondary device can potentially reduce adoption and increase account lockouts \cite{doerflerEvaluatingLoginChallenges2019, decristofaroComparativeUsabilityStudy2014}.

\section{Previous Research and Our Contribution}
\label{sec:prevresearch}

Prior research has investigated the efficacy of multi-factor authentication (MFA) challenges for consumer accounts, such as the Microsoft Account (MSA) and the Google Account, and found that 1) MFA challenges are highly effective in preventing account compromise, 2) some types of additional authentication forms are more effective than others at preventing account compromise, and 3) there are trade-offs between prevention effectiveness, ease of adoption, and ease of use \cite{doerflerEvaluatingLoginChallenges2019,adamsUsersAreNot1999,ulqinaku2FAPP2ndFactor2019,bonneauQuestReplacePasswords2012}. Recently, the continued effectiveness of MFA has been called into question \cite{MultifactorAuthenticationLess, HasMultiFactorAuthentication}.

Consumer accounts are pervasive and primarily grant access to free services, including personal email, media personalization, and instant messaging. In contrast, accounts provided by enterprises and governmental institutions to their workforce and customers often grant access to different types of data and resources, such as payment information, servers containing aggregated financial data, and computational resources. These commercial accounts often rely on protection from commercial identity products like Microsoft's Azure Active Directory (AAD) and Okta's Workforce Identity Cloud, although some large providers, such as Amazon, use their own in-house solutions \cite{liveretosCustomerIdentityAccess2022}. 

During our measurement period, commercial accounts constituted approximately one-third of the total accounts in use within a given month. Unlike consumer account users, who directly register with authentication providers, commercial account users must register with an intermediate layer known as the tenant administrator, typically their own institution. For instance, a university professor's account is provided by the university itself, even when the authentication services are ultimately performed by an identity provider like Microsoft. The tenant administrator, the university in our example, is responsible for registering and maintaining accounts, defining security policies, including which resources will require MFA and the type of MFA to be used, and delivering first-level support\cite{dawoudZeroTrustDeployment2020}.

Although consumer account data may occasionally hold value, gaining access to commercial accounts is generally more valuable\cite{castellMitigatingOnlineAccount}. Consequently, bad actors may dedicate more time and resources targeting commercial accounts, which may result in MFA having different effectiveness for commercial accounts. This paper focuses on evaluating the effectiveness of security solutions applied to commercial accounts and comparing these findings to previous research conducted on consumer accounts.

\section{Methodology and Data}

Our goal is to determine the effectiveness of MFA in preventing account compromise in the population of commercial accounts. It is generally not possible for an authentication provider to obtain the exact number of account compromises in a population without resorting to sampling and manual reviews. When users detect an account compromise, they may simply change their passwords and not notify their administrators. Even when the administrators are notified, they may choose not to notify the authentication provider. Therefore, methods that rely on the authentication provider using reported account compromises will result in an undercount of the actual rate. On the other hand, it is cost-prohibitive for an authentication provider that has billions of accounts to manually review all suspected compromises. Therefore, we have to rely on sampling methods. 

To achieve our goal, we obtained a list of active Microsoft Azure Active Directory users that had their account reviewed due to suspicious activity between April 22, 2022, and September 22, 2022. Some accounts had MFA configured, and some did not. If the account had suspicious activity and had MFA configured, a challenge was automatically issued. A sample of the sessions was retroactively reviewed by a specialized team that examines account logs and determines whether a compromise occurred or not. If a compromise was detected, the account was sanitized, and the user notified.  

To estimate the proportion of compromised accounts in the whole population, we use the benchmark multiplier method\cite{hickmanIndirectMethodsEstimate2005}, commonly used in epidemiological research in situations where individuals tend to underreport the actual frequency of an event. The benchmark multiplier method requires two datasets: one, the benchmark, has a complete and accurate count of the event being studied for a subgroup of the population. The other dataset is a representative sample from the population, used to estimate the proportion of the population represented by the benchmark. The reciprocal of that proportion is called the multiplier.  

In our case, the benchmark is the set of accounts that were manually reviewed by the account specialists. For this dataset, we have the exact numbers of accounts compromised. Our benchmark is divided into two MFA categories (MFA enabled and MFA not enabled). To connect the benchmark with the total population, we obtain a random sample of accounts of the whole population for each category and calculate the proportion $\pi$ of accounts that are in our benchmark and have been compromised. 

Using the methodology laid out in \cite{mojtabaiEstimatingPrevalenceSubstance2022}, given a benchmark of size $\hat{N}_x$ and the probability $\hat{\pi}$ for members of the representative sample to be in the benchmark, we can estimate $\hat{N}_y$, the number of accounts compromised in the population as 

\[
\hat{N}_y = \frac{\hat{N}_x}{\hat{\pi}}
\]

For each category, following \cite{bollaertsImprovedBenchmarkmultiplierMethod2013}, the proportion $\pi$ is distributed 
$\hat{\pi} \sim \beta(x+1, n+x+1)$, where $n$ is the size of the representative sample and $x$ is the number of members of that sample that share the benchmark's characteristics. In addition, even if $\hat{N}_x$ and $\hat{\pi}$ are unbiased, $\hat{N}_y$ is a biased estimator of $N_y$ because of its non-linearity with respect to $\hat{\pi}$. Therefore, following \cite{bollaertsImprovedBenchmarkmultiplierMethod2013}, we use a bias-corrected estimator:

\[
\hat{N}_y = \frac{\hat{N}_x}{\hat{\pi}} - \frac{1}{n} \hat{N}_x \frac{(1-\hat{\pi})}{\hat{\pi}}
\]

We use a Monte Carlo simulation to estimate $\hat{N}_y$ for each category. We run each simulation 1,000 times. Our 95\% confidence intervals are based on the 2.5\% and 97.5\% percentiles of the 1,000 simulated estimates. The estimates for the proportion $\hat{\pi}$ are in Table \ref{tab:table1}.

\section{Results}

Our results are shown in Table \ref{tab:table1}, where (a) is the number of compromises measured in the benchmark, (b) is the median number of compromises estimated in the population, and (c) is the median percentage number of compromises estimated in the population. 

According to these estimates, the median estimated compromise rate of MFA accounts is 0.0079\%, which means that MFA accounts have a protection factor better than 99.99\% for commercial accounts, in line with estimates previously found for consumer accounts.

\begin{table}[H]

 \caption{Results with and without MFA}
  \centering
  \begin{tabular}{lrrrr}
    \toprule
    Category     & $\hat{\pi}$ (95\% CI)  & (a) & (b) & (c) \\
    \midrule
    With MFA & 2.20\% - 3.01\% & 1,525 & 59,414 & 0.0079\% \\
    Without MFA & 0.18\% - 0.26\% & 15,195 & 7,085,925 & 1.0071\% \\
    \bottomrule
  \end{tabular}
  \label{tab:table1}
\end{table}

We also calculate effectiveness as the proportion of risk reduction, using the same formula used to calculate vaccine effectiveness. A member of the population treated with MFA has an estimated median risk of 0.0079\%, while a member of the population not treated with MFA has an estimated median risk of 1.0071\%. Therefore, the risk reduction of using MFA is

\[
\text{RR} = 1-\frac{\text{treatment}}{\text{no treatment}} = 1 - \frac{0.0079\%}{1.0071\%} = 99.22\%
\]

Another way of measuring the effectiveness of MFA is calculating the ratio of account compromises coming from accounts with and without MFA enabled, as shared by Microsoft in the 2020 RSA conference \cite{Microsoft99Compromised}. Using the median estimate of compromises in our data, we find that $1-\frac{59,414}{(7,085,925+59,414)} = 99.17\%$ of the compromised accounts did not have MFA enabled. This is slightly lower than the number found in 2019 by \cite{Microsoft99Compromised}. However, between 2019 and 2022, we have observed the adoption of MFA to have increased by over 400\%.

\section{Accounts with Known Leaked Credentials}

In 2019, Google published a study about consumer accounts that found that challenges and MFA prevented 100\% of automated attacks, 96\% of bulk phishing attacks, and 76\% of targeted attacks \cite{doerflerEvaluatingLoginChallenges2019}. These percentages were calculated for a subset of accounts for which an attack was known to have happened, and therefore not directly comparable to our figures above. 

We obtained a sample of 128,000 accounts that had their passwords leaked between April and September of 2022. Users were immediately notified. We retroactively studied those accounts for 30 days prior to the discovery of the credential leak. Reviewing the accounts manually, we found 7,861 accounts that had MFA and for which we could confirm that attackers used the passwords to try to obtain access to protected resources. For those accounts, we found that MFA prevented 98.6\% of the attacks. 

For this sample, we were able to analyze the specific type of MFA used and its performance. The detailed results are in Table \ref{tab:table2}. Similar to \cite{doerflerEvaluatingLoginChallenges2019}, we find that SMS was 40.8\% less effective than Microsoft Authenticator, a mobile application specifically designed for multi-factor authentication. 

\begin{table}[H]

 \caption{Results with and without MFA}
  \centering
  \begin{tabular}{lr}
    \toprule
    MFA Type & Failure Rate \\
    \midrule
    Authenticator OTP & 0.99\% \\
    Authenticator Notifications & 0.97\% \\
    SMS & 1.66\% \\
    \midrule
    \textbf{Total} & \textbf{1.44\%} \\
    
    \bottomrule
  \end{tabular}
  \label{tab:table2}
\end{table}

\section{Conclusion}

In this research, we have conducted the first analysis of the effectiveness of multifactor authentication (MFA) in securing commercial accounts. By leveraging the benchmark-multiplier method and manually reviewing a sample of potentially compromised accounts, we have found that 99.99\% of accounts with MFA enabled remained protected throughout the investigation period. Our findings further demonstrate that implementing MFA leads to a 99.22\% reduction in the risk of compromise across the entire population, and a 98.56\% reduction even in cases where credentials have been leaked. These results for commercial accounts are similar to the results reported in previous studies for consumer accounts.

In addition, our study finds that dedicated MFA applications outperform SMS-based authentication, although both methods are significantly more effective than not employing MFA at all. In light of these findings, we strongly advocate for the default activation of MFA in commercial accounts to bolster cybersecurity measures, as already required by many institutions\cite{pomputiusReviewTwoFactorAuthentication2018a}.

%Bibliography
\bibliographystyle{unsrt}  
\bibliography{references}  

\begin{thebibliography}{10}

\bibitem{freemanWhoAreYou2016}
David~Mandell Freeman, Sakshi Jain, Markus Dürmuth, Battista Biggio, and
  Giorgio Giacinto.
\newblock Who {{Are You}}? {{A Statistical Approach}} to {{Measuring User
  Authenticity}}.
\newblock {\em NDSS}, 16:21--24, 2016.

\bibitem{ogormanComparingPasswordsTokens2003}
L.~O'Gorman.
\newblock Comparing passwords, tokens, and biometrics for user authentication.
\newblock {\em url =
  {https://slate.com/technology/2022/02/google-multifactor-authentication-effective-research.html},ceedings
  of the IEEE}, 91(12):2021--2040, 2003.

\bibitem{liveretosCustomerIdentityAccess2022}
Anastasios Liveretos and Ivo Draganov.
\newblock Customer {{Identity}} and {{Access Management}} ({{CIAM}}): An
  {{Overview}} of the {{Main Technology Vendors}}.
\newblock {\em International Journal of Economics and Management Systems}, 07,
  2022.

\bibitem{dasWhyJohnnyDoesn2018a}
Sanchari Das, Andrew Dingman, and L.~Jean Camp.
\newblock Why {{Johnny Doesn}}’t {{Use Two Factor A Two-Phase Usability
  Study}} of the {{FIDO U2F Security Key}}.
\newblock In Sarah Meiklejohn and Kazue Sako, editors, {\em Financial
  {{Cryptography}} and {{Data Security}}}, volume 10957, pages 160--179.
  {Springer Berlin Heidelberg}, 2018.

\bibitem{doerflerEvaluatingLoginChallenges2019}
Periwinkle Doerfler, Kurt Thomas, Maija Marincenko, Juri Ranieri, Yu~Jiang,
  Angelika Moscicki, and Damon McCoy.
\newblock Evaluating {{Login Challenges}} as a {{Defense Against Account
  Takeover}}.
\newblock In {\em The {{World Wide Web Conference}} on - {{WWW}} '19}, pages
  372--382. {ACM Press}, 2019.

\bibitem{decristofaroComparativeUsabilityStudy2014}
Emiliano De~Cristofaro, Honglu Du, Julien Freudiger, and Greg Norcie.
\newblock A {{Comparative Usability Study}} of {{Two-Factor Authentication}}.
\newblock http://arxiv.org/abs/1309.5344, 2014.

\bibitem{adamsUsersAreNot1999}
Anne Adams and Angela Sasse.
\newblock Users {{Are Not}} the {{Enemy}}.
\newblock {\em Communications of the ACM}, 42:40--46, 1999.

\bibitem{ulqinaku2FAPP2ndFactor2019}
Enis Ulqinaku, Daniele Lain, and Srdjan Capkun.
\newblock {{2FA-PP}}: 2nd factor phishing prevention.
\newblock In {\em Proceedings of the 12th {{Conference}} on {{Security}} and
  {{Privacy}} in {{Wireless}} and {{Mobile Networks}}}, pages 60--70. {ACM},
  2019.

\bibitem{bonneauQuestReplacePasswords2012}
Joseph Bonneau, Cormac Herley, Paul Oorschot, and Frank Stajano.
\newblock The {{Quest}} to {{Replace Passwords}}: {{A Framework}} for
  {{Comparative Evaluation}} of {{Web Authentication Schemes}}.
\newblock In {\em 2012 {{IEEE Symposium}} on {{Security}} and {{Privacy}}},
  pages 553--567, 2012.

\bibitem{MultifactorAuthenticationLess}
Is multifactor authentication less effective than it used to be?
\newblock
  https://slate.com/technology/2022/02/google-multifactor-authentication-effective-research.html,
  2022.

\bibitem{HasMultiFactorAuthentication}
Has {{Multi-Factor Authentication Failed Us}}?
\newblock https://www.pcmag.com/news/has-multi-factor-authentication-failed-us,
  2023.

\bibitem{dawoudZeroTrustDeployment2020}
Tarek Dawoud.
\newblock Zero {{Trust Deployment Guide}} for {{Microsoft Azure Active
  Directory}}.
\newblock
  https://www.microsoft.com/en-us/security/blog/2020/04/30/zero-trust-deployment-guide-azure-active-directory/,
  2020.

\bibitem{castellMitigatingOnlineAccount}
Michelle Castell.
\newblock Mitigating {{Online Account Takeovers}}: {{The Case}} for
  {{Education}}.
\newblock In {\em Retail Payments Risk Forum}, 2013.

\bibitem{hickmanIndirectMethodsEstimate2005}
Matthew Hickman and Colin Taylor.
\newblock Indirect {{Methods}} to {{Estimate Prevalence}}.
\newblock In Zili Sloboda, editor, {\em Epidemiology of {{Drug Abuse}}}, pages
  113--131. {Springer-Verlag}, 2005.

\bibitem{mojtabaiEstimatingPrevalenceSubstance2022}
Ramin Mojtabai.
\newblock Estimating the {{Prevalence}} of {{Substance Use Disorders}} in the
  {{US Using}} the {{Benchmark Multiplier Method}}.
\newblock {\em JAMA Psychiatry}, 79(11):1074, 2022.

\bibitem{bollaertsImprovedBenchmarkmultiplierMethod2013}
Kaatje Bollaerts, Marc Aerts, and Andre Sasse.
\newblock Improved benchmark-multiplier method to estimate the prevalence of
  ever-injecting drug use in {{Belgium}}, 2000–10.
\newblock {\em Archives of Public Health}, 71(1):10, 2013.

\bibitem{Microsoft99Compromised}
Microsoft: 99.9\% of compromised accounts did not use multi-factor
  authentication.
\newblock
  https://www.zdnet.com/article/microsoft-99-9-of-compromised-accounts-did-not-use-multi-factor-authentication/,
  2020.

\bibitem{pomputiusReviewTwoFactorAuthentication2018a}
Ariel~F. Pomputius.
\newblock A {{Review}} of {{Two-Factor Authentication}}: {{Suggested Security
  Effort Moves}} to {{Mandatory}}.
\newblock {\em Medical Reference Services Quarterly}, 37(4):397--402, 2018.

\end{thebibliography}

\end{document}